\documentclass[twocolumn,aps]{revtex4}
\usepackage{graphicx,epsf}

\begin{document}

\title{ Stochastic Potential Switching Algorithm for Monte Carlo 
Simulations of Complex Systems }

\author{C.H.~Mak}

\affiliation{
  Department of Chemistry, 
  University of Southern California, \\
  Los Angeles, California 90089-0482, USA }

\date{\today}


\begin{abstract} 

This paper describes a new Monte Carlo method based on a novel stochastic 
potential switching algorithm.  This algorithm enables the equilibrium 
properties of a system with potential $V$ to be computed using a 
Monte Carlo simulation for a system with a possibly less complex stochastically 
altered potential $\tilde V$.  By proper choices of the stochastic switching 
and transition probabilities, it is shown that detailed balance can be strictly 
maintained with respect to the original potential $V$.  The validity of
the method is illustrated with a simple one-dimensional example.  
The method is then generalized to multidimensional systems with any 
additive potential, providing a framework for the design of more 
efficient algorithms to simulate complex systems.  
A near-critical Lennard-Jones fluid with more than 20000 particles 
is used to illustrate the method.  The new algorithm produced a much 
smaller dynamic scaling exponent compared to the Metropolis method 
and improved sampling efficiency by over an order of magnitude.

\end{abstract}

\maketitle

\section{Introduction} \label{sect:introduction}

Simulations of complex molecular systems are generally carried out using 
either the molecular dynamics (MD) or the Monte Carlo (MC) method.
Each method has its own merits.
The MD method \cite{87alla, 57ald1208}, 
based on the integration of the classical 
equations of motion of the particles, is conceptually the simpler of the 
two.  With currently available computer power, MD simulations 
generally cannot be carried out for 
very long time scales for very large systems, making the extraction of true 
equilibrium properties often difficult.  
The MC method \cite{77vala, 86bina}, on the other hand, 
relies on stochastic dynamics to generate members of the desired ensemble.   
MC has the ability to execute non-physical large-scale transitions
that are impossible in MD and has the potential to reach 
equilibrium much faster.  However, devising these large-scale transitions
that have reasonable acceptance probabilities is not always straightforward.

To search for ways to enable large-scale transitions to be carried out with 
higher probability in MC simulations, one must tackle the core problem, 
which is the complexity of the interactions among 
the particles in the system.  If there were no interactions among these 
particles, any transition, regardless of its scale, would always be accepted.  
When a move is made in MC, the interactions involving those particles that
are being moved will change.  In general, the larger the scale of the move and 
the more complicated the interactions are, the larger the change in the   
potential becomes.  Consequently, large-scale MC moves are very unlikely to 
be accepted. 

A natural questions arises:  Is it possible to reduce the complexity of
the interactions among the particles, for instance, by replacing the
actual potential $V$ by a less complex potential $\tilde V $?  
One  possibility is proposed in this paper.  With this method, 
it is indeed possible 
to replace the original potential $V$ by an {\em arbitrary} $\tilde V$. 
But the procedure has to follow 
a carefully constructed algorithm to guarantee that detailed balance 
with respect to the original potential is maintained, 
so that the correct statistics are produced.
An idea similar to this has been exploited in 
a number of previously proposed Monte Carlo methods, 
such as J-walking \cite{90fra2769, 92fra5713, 98cur1643}, 
simulated tempering \cite{92mar451, 96tes155}, 
parallel tempering \cite{99fal1754, 99yan9509, 00yan1276, 02fal5419}, 
catalytic tempering \cite{00sto11164}, 
multicanonical J-walking \cite{99xu10299} 
and the approximate potential method \cite{03gel7747}.  
But we will show that when generalized to multidimensional systems, 
the present method provides flexibilities and potential advantages that are not 
available with these previous methods and establishes a theoretical framework 
for the design of possibly more efficient algorithms for 
simulating complex systems.

\section{The SPS Idea} \label{sect:algorithm}

Let $x$ be the configuration of a $N$-dimensional system and $V(x)$ the 
potential energy divided by the Boltzmann constant $k_B$.  The
statistical weight of each member of 
the canonical ensemble is $\exp(-V(x)/T)$, $T$ being the absolute 
temperature.  A MC algorithm that generates configurations 
consistent with their statistical weights 
can be constructed from any set of transition 
rules, as long as the transition 
probabilities $W$ between every pair $x$ and $x^\prime$ 
satisfy the detailed balance condition: 
\begin{equation}
e^{-V(x)/T} W(x \to x^\prime) 
= e^{-V(x^\prime)/T} W(x^\prime \to x). 
\label{eq:detailed}
\end{equation}

\begin{figure}
\includegraphics[width=0.8\columnwidth]{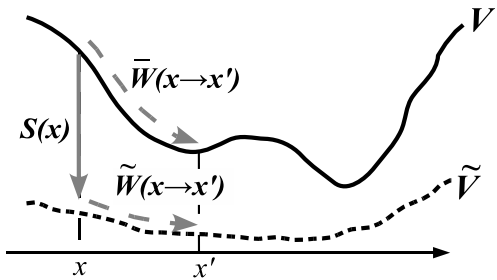}
\caption[]{
\label{fig:SPS}
Illustration of the SPS procedure.  (See text for details.)
}
\end{figure}

The new MC algorithm we propose proceeds as follows:
\begin{enumerate}
\item
First, consider changing the system potential $V$ to an arbitrary potential 
$\tilde V$.  This ``potential switching'' decision is carried out with a 
stochastic switching probability 
\begin{equation}
S(x) = e^{(\Delta V(x) - \Delta V^*)/T}, \label{eq:S}
\end{equation}
with $\Delta V(x) = V(x) - \tilde V(x)$ and $\Delta V^*$ is a constant 
greater than or equal to the 
maximum value of $\Delta V(x)$ over {\em all} $x$.
By incorporating 
$\Delta V^*$ into Eqn.(\ref{eq:S}), we ensure that $S(x)$ is between
0 and 1.
Note that if $\Delta V^*$ is very large, the resulting $S(x)$ will 
be small, making switching very infrequent.  
For systems with a $V$ that is bounded from above, it is always possible 
to choose a $\tilde V$ which yields a finite $\Delta V^*$.  
Systems with an unbounded $V$ will be addressed in 
Sect.~\ref{sect:generalization}.
\item
If the switch is made, the configuration of the system is 
moved from $x$ to $x^\prime$ with transition probability 
$\tilde W(x \to x^\prime)$ chosen to 
satisfy detailed balance on the 
{\em switched potential} $\tilde V$, i.e.
\begin{equation}
e^{-\tilde V(x)/T} \tilde W(x \to x^\prime) 
= e^{-\tilde V(x^\prime)/T} \tilde W(x^\prime \to x). \label{eq:tildeW}
\end{equation}
Any $\tilde W$, such as Metropolis \cite{53met1087}, may be used here 
as long as it satisfies Eqn.~(\ref{eq:tildeW}).
\item
If the switch is unsuccessful, the configuration of
the system is moved from $x$ to $x^\prime$ with transition probability
$\bar W(x \to x^\prime)$ chosen to satisfy detailed balance on a 
{\em pseudopotential} 
\begin{equation}
\bar V(x) = V(x) - T \ln [1 - S(x)], \label{eq:Vbar}
\end{equation}
Similar to step 2, any $\bar W$ may be used here 
as long as it satisfies detailed balance on $\bar V$.
\end{enumerate}
After the move from $x \to x^\prime$ is made (accepted or rejected), the 
cycle is over and the algorithm returns to step 1.
This stochastic potential switching (SPS) idea, 
with the relevant potential switching and subsequent 
MC transition probabilities, is illustrated schematically in 
Fig.~\ref{fig:SPS}.  
Obviously, the SPS algorithm can be used alone in a simulation 
(if the moves are ergodic) or mixed with other MC moves.

With the algorithm defined above, 
it is easy to prove that the composite transition probability:
\begin{equation}
W(x \to x^\prime) = S(x) \tilde W(x \to x^\prime) 
+ [1 - S(x)] \bar W(x \to x^\prime), \label{eq:W}
\end{equation}
when substituted into Eqn.(\ref{eq:detailed}), indeed satisfies 
detailed balance with respect to the original potential $V$.  
Therefore, the MC trajectory generated by this 
SPS idea will produce a sequence of configurations $\{x\}$ that is 
consistent with the canonical ensemble for a system with potential $V(x)$.
It is important to emphasize that the choice of $\tilde V$ is completely
arbitrary --- the proof works for all $\tilde V$.  In a real application,  
one can exploit this arbitrariness to select a 
$\tilde V$ that may be either less complex than the original $V$ or 
less costly to compute.
For a system where the potential is a sum of additive 
terms $V = \sum_i V_i$, 
often the case for many-particle systems, 
the switching decision can be applied to each $V_i$ 
separately.  This generalization will be described in 
Sect.~\ref{sect:generalization}.

In form described above, the SPS idea is conceptually related to 
J-walking \cite{90fra2769, 92fra5713, 98cur1643}, 
parallel tempering \cite{99fal1754, 99yan9509, 00yan1276, 02fal5419},  
and the ``approximate potential'' method \cite{03gel7747}.  
In J-walking, the simulation is stochastically switched to a 
configuration sampled from a higher-temperature ($T^\prime$) simulation 
of the same potential with properly chosen transition probabilities.  
This is essentially the same as using a potential that is attenuated by 
a factor $T/T^\prime$ as $\tilde V$ in SPS.  Similarly, in parallel 
tempering, the exchange of replicas between two different temperatures 
is equivalent to having one switched to an attenuated potential and the 
other to a higher potential.
In the approximate potential method, the simulation is switched to 
an approximate potential, and the new configuration produced on the
approximate potential is accepted or reject at the end with 
a ``correction'' rate that is designed to 
maintain detailed balance with respect to the original potential; 
whereas in SPS, 
the switching decision is made {\em before} the move, so that 
the subsequent update on $\tilde V$ will always be accepted.
Even though SPS is conceptually akin to these other methods, we will show 
in Sect.~\ref{sect:generalization} that 
when SPS is generalized to multidimensional systems, its offers 
flexibilities and potential advantages that are not currently 
available in these related methods.

\section{Example: A one-dimension model} \label{sect:example}

We use a simple one-dimensional example to illustrate the 
basic SPS idea.  
The model we selected was a 
harmonic potential $V(x) = \frac{1}{2} x^2$, 
with $x$ confined to within the range $[-1,1]$.  We used 
five different $\tilde V(x) = \frac{1}{2}x^n$, 
with $n$ = 0, 1, 2, 3 and 4 to demonstrate 
that the ensemble averages were indeed invariant with the choice of 
$\tilde V$ and the choice of $\tilde V$ is hence arbitrary.  
For each of the five $\tilde V$, we sampled $x$ 
according to the SPS algorithm
using simple Metropolis moves on both $\tilde V$ and $\bar V$.  
$n=2$ is a special case, because for $n=2$, 
$\tilde V = V$, so the switch is make with unit 
probability.  The SPS algorithm for $n=2$ 
is thus equivalent to the normal Metropolis (i.e. non-SPS) algorithm.

The moments $\langle x^m \rangle$ at $T = 0.2$ 
for $m$ = 1 to 10 are shown in Table~\ref{tab:example} for the five 
different $\tilde V$.  The results are clearly invariant with the 
choice of $\tilde V$ to within statistical errors, showing that detailed
balance is strictly satisfied with respect to $V$ for any $\tilde V$.
The switching rate $R_S$ is also shown for each $\tilde V$.  $R_S$
is in general lower for those $\tilde V$ (particularly 
$n$ = 1 and 3) that are very different from the original potential $V$.  
Notice also that $n$ = 0 corresponds to a flat potential.  In this case, 
if the switch is made, the potential is completely turned off.

\begin{table*}
\caption[]{
\label{tab:example}
MC results for the model system $V(x) = \frac{1}{2} x^2$ at $T=0.2$ using 
5 different $\tilde V = \frac{1}{2} x^n$.  $\langle x^m \rangle$ are 
the $m$-th moments measured by SPS-MC.  $R_S$ are the observed switching 
rate for each $\tilde V$.  The uncertainty of the last digit is shown 
in parentheses.
}
\begin{tabular}{| c | r r r r r |} \hline
                       &$n=0$     &$n=1$     &$n=2$     &$n=3$     &$n=4$     \\ \hline
$\langle x^1\rangle$   &$-0.0003(8)$&$-0.0009(8)$&$ 0.0008(8)$&$ 0.0004(8)$&$ 0.0010(8)$ \\  
$\langle x^2\rangle$   &$ 0.1698(4)$&$-0.1698(4)$&$ 0.1702(4)$&$ 0.1701(4)$&$ 0.1701(4)$ \\
$\langle x^3\rangle$   &$-0.0001(4)$&$-0.0005(4)$&$-0.0001(4)$&$ 0.0001(4)$&$ 0.0003(4)$ \\
$\langle x^4\rangle$   &$ 0.0716(3)$&$ 0.0723(3)$&$ 0.0720(3)$&$ 0.0720(3)$&$ 0.0721(3)$ \\
$\langle x^5\rangle$   &$-0.0000(3)$&$-0.0004(3)$&$-0.0001(3)$&$ 0.0000(3)$&$ 0.0001(3)$ \\
$\langle x^6\rangle$   &$ 0.0416(2)$&$ 0.0424(2)$&$ 0.0418(2)$&$ 0.0419(2)$&$ 0.0420(2)$ \\
$\langle x^7\rangle$   &$-0.0000(2)$&$-0.0003(2)$&$-0.0001(2)$&$-0.0000(2)$&$ 0.0000(2)$ \\
$\langle x^8\rangle$   &$ 0.0283(2)$&$ 0.0291(2)$&$ 0.0285(2)$&$ 0.0286(2)$&$ 0.0287(2)$ \\
$\langle x^9\rangle$   &$-0.0000(2)$&$-0.0002(2)$&$-0.0001(2)$&$-0.0000(2)$&$-0.0000(2)$ \\
$\langle x^{10}\rangle$&$ 0.0211(2)$&$ 0.0218(2)$&$ 0.0213(2)$&$ 0.0214(2)$&$ 0.0215(2)$ \\ \hline
$\Delta V^*$        & 0.5      & 1.       & 0.       & 1.       & 0.125     \\ 
$R_S$               & 0.151    & 0.030    & 1.000    & 0.020    & 0.699     \\ \hline
\end{tabular}
\end{table*}

\section{Relationship to Ising-Type Cluster Algorithms} \label{sect:swendsen}

A MC method that is capable of executing large-scale moves with high
probability was proposed by Swendsen and 
Wang \cite{87swe86} for the Ising model in 1987.  This method later led to the 
discovery of a class of methods now collectively known as ``cluster 
algorithms'' \cite{02blo58}.  These cluster algorithms 
can be shown to be special cases of the SPS algorithm.  
Whereas these cluster algorithms permit large-scale MC moves, they 
are largely restricted to discrete models and are not generally applicable to 
molecular simulations.
The SPS method provides a way to transfer the ideas behind these 
cluster algorithms to continuous systems.

In the original Swendsen-Wang algorithm, the interaction between each 
pair of Ising spins $\sigma_i$ and $\sigma_j$ is stochastically deleted 
with a probability $p_d = \exp[-J(\sigma_i\sigma_j - 1)]$, where $J$ is the
Ising interaction.  When an interaction is successfully deleted, it has no
effect on the subsequent MC move.  On the other hand, if an interaction 
is not deleted, it is frozen such that the value of this interaction is 
constrained to remain constant in the subsequent MC move.  After all 
the interactions have been either deleted or frozen, the spins break up
into clusters of spins having frozen interactions.  
Each cluster can be flipped independently of the others.  
Swendsen and Wang showed that this cluster
algorithm satisfies detailed balance and it produces much faster 
equilibration compared to the conventional Metropolis algorithm 
\cite{53met1087}.

The Ising model has potential $V = \sum_{(ij)} V_{ij}$, 
where $V_{ij} = -J\sigma_i\sigma_j$ and the sum goes over all 
nearest-neighbor pairs.   It is easy to show that the Swendsen-Wang 
algorithm can be derived from the SPS algorithm by making the 
special choice $\tilde V_{ij} = 0$.  As such, the SPS algorithm 
can be considered as a ``generalized'' cluster algorithm.  However, 
we choose not to use this terminology because calling SPS a generalized 
cluster algorithm would improperly imply some geometric origin.  Whereas 
in discrete systems such as the Ising model there is an obvious geometric 
interpretation for the SPS algorithm, there may not be any in more general 
continuous systems.

\section{Generalization to Systems with Additive Potentials} 
\label{sect:generalization}

While the validity of the SPS algorithm is clear for the basic 
case considered in Sect.~\ref{sect:algorithm}, 
the utility of the SPS algorithm in this form is rather limited.   
There are several reasons why this formulation of the SPS 
algorithm may not be very practical.   
(1) The choice of a good $\tilde V$ is not obvious.  
(2) Because $S(x)$ in Eqn.~\ref{eq:S} 
is scaled by $e^{-\Delta V^*/T}$, unless $\tilde V$ is close to $V$ 
{\em everywhere}, the switching frequency will be in general small.
(3) For a $V$ that is not bounded from above (as in systems with 
repulsive interactions), there may not be a way to choose a $\tilde V$ 
that keeps $\Delta V^*$ finite.

To make the SPS algorithm more useful, we must first generalize it to 
an additive potential $V$ that can be written as a sum of two or more terms.
Any $V$ can be decomposed into an arbitrary sum.  
Some systems, such as those with pairwise interactions, have potentials 
that break up naturally into a sum of terms.  In other situations, the 
potential may have two or more distinct parts that are responsible 
for different physical phenomena, such as the repulsive and attractive 
part of a Lennard-Jones potential \cite{83wee787}.  
We will see that the usefulness of the SPS algorithm is related to 
how $V$ is decomposed.
Our formulation here is inspired by the ideas of Kandel {\em et al.}
\cite{88kan1591} who have provided a generalization of the Swendsen-Wang 
algorithm \cite{87swe86} for discrete-state (Ising) models.  

To illustrate the generalization of the SPS 
algorithm to a continuous system with an additive potential, we consider
a potential with just two terms $V(x) = V_1(x) + V_2(x)$.  
Extension to more than two terms is straightforward.  
We can apply the SPS algorithm in 
Sect.~\ref{sect:algorithm} to each of the terms separately, 
attempting to switch $V_1$ to a new $\tilde V_1$ and $V_2$ to 
another $\tilde V_2$ with switching probabilities 
$S_1(x) = e^{(\Delta V_1(x) - \Delta V_1^*)/T}$ and
$S_2(x) = e^{(\Delta V_2(x) - \Delta V_2^*)/T}$, 
respectively.  Since the potential terms are additive, the 
switching of $V_1$ and $V_2$ are independent of each other and can
be performed in any order.

For two terms in the potential, 
there are four possible outcomes of the switching decision.  For 
each one, the next part of the simulation will proceed on a different 
potential:  
(1) if both $V_1$ and $V_2$ are switched, the new
potential becomes $\tilde V_1 + \tilde V_2$;
(2) if both $V_1$ and $V_2$ are not switched, the new
potential becomes $\bar V_1 + \bar V_2$;
(3) if $V_1$ is switched but $V_2$ is not, the new
potential becomes $\tilde V_1 + \bar V_2$; 
(4) if $V_1$ is not switched but $V_2$ is, the new
potential becomes $\bar V_1 + \tilde V_2$, where 
$\bar V_1$ and $\bar V_2$ are defined as in Eqn.~\ref{eq:Vbar}.
After the switch is made, the system can be moved from 
configuration $x \to x^\prime$ on the new potential.  
At the end of the move, the original potential
can be restored to restart the switching process anew.
One can easily show that with this algorithm, detailed balance with 
respect to the original potential is strictly obeyed along 
{\em any one} of the four pathways.  This is a direct result of the 
additivity of $V_1$ and $V_2$.  The sum over all four 
pathways therefore also obeys detailed balance.

In the above, 
we have considered switching both terms in $V$, 
but this needs not be.  In fact, we can 
apply the switching to an arbitrary subset of terms.  For example, we 
may consider switching only $V_1$ to $\tilde V_1$ and keeping $V_2$ 
``alive''.  If the switching is successful, the new potential becomes 
$\tilde V_1 + V_2$; otherwise, it is $\bar V_1 + V_2$.
This scenario is equivalent to using a $\tilde V_2 = V_2$ and thus 
also satisfies detailed balance.
This strategy may be useful, for example, in the case of a $V$ that is 
not bounded from above.
In this case, we can decompose the potential into the repulsive 
(unbounded) and attractive (bounded) parts, but apply the switching 
only to the attractive part.

It should now be clear why decomposing $V$ into many additive terms 
makes the SPS algorithm more practical.  In the original formulation of 
the SPS algorithm in 
Sect.~\ref{sect:algorithm}, 
the entire $V = \sum_\ell V_\ell$ is switched to 
$\tilde V$.  The probability for the simultaneous 
switching of {\em all} $V_\ell$ is $\prod_\ell S_\ell$.  If the potential 
contains a large number of terms, the total switching probability will be 
small even if each individual $S_\ell$ is close to unity.  Therefore, 
switching the entire $V$ is almost impossible, but 
individual terms in $V$ can be switched with a much higher probability.  

Coupled with good physical insights, the additivity of $V$ can be 
exploited to devise efficient SPS algorithms that may be more optimal 
than others.  Some systems, such as those with pairwise interactions, 
have a natural decomposition for $V$.  This may be used to guide the 
search for an optimal breakup.  On the other hand, there may be a 
totally unphysical breakup that affords higher efficiency.  The 
additivity of $V$ offers immense possibilities.
In the next section, we will illustrate this using a nontrivial 
many-particle example.

In fact, crude elements of the basic SPS idea have already appeared in 
one of our recent studies on the Monte Carlo simulations of imaginary-time 
path integrals \cite{04mak6760}, and these ideas have been proven useful 
for accelerating the sampling of stiff paths.  
The strategy proposed there was later implemented in a large-scale 
path integral simulation of superfluid molecular H$_2$ clusters
\cite{05makxxxxa}.
These studies motivated us to refine the crude ideas contained in 
those two papers and formulate the more general theoretical framework for 
the SPS algorithm that has been presented in this section.

\section{Example: A Lennard-Jones fluid near its critical point} 
\label{sect:lj}

The correlation length of a system diverges near the critical point.  
Small local fluctuations of the system at large 
separations become correlated with each other, making 
Monte Carlo simulations extremely sluggish \cite{86bina}.  
This so-called ``critical slowing-down'' problem leads to 
extremely long equilibration time for Monte Carlo simulations 
that employ only local updates, such as the conventional Metropolis 
algorithm.  Nonlocal moves can also be made using the 
Metropolis algorithm, but such moves are almost always rejected 
because of the reason given in the last section.

We will demonstrate how the SPS method can be used to deal with 
the critical slowing-down problem in a Lennard-Jones fluid.  The 
simulations were carried out in a cubic box with periodic boundary 
condition.  A Lennard-Jones potential 
$u(r) = 4\epsilon \left[ (\sigma/r)^{12} - (\sigma/r)^{6} \right]$ 
that is truncated but unshifted at $r_c = 2.5 \sigma$ was used 
for the calculations.  
Previously, the critical temperature $T_c$ 
and density $\rho_c$ for this system were found to be 
$k_B T_c = 1.1853 \epsilon$ and $\rho_c \sigma^3 = 0.3197$ \cite{95wil602}.  
To check scaling, we vary the box length $L$ from $L/\sigma$ = 10 to 40, 
using up to 20464 particles to maintain a fixed density.
Since the heat capacity is expected to diverge as $\vert T - T_c\vert^\alpha$
\cite{76maa}, the slowing-down problem should be manifested in the 
energy measurement.  We compared the scaling of the autocorrelation time 
for the energy estimator with the box length $L$ of the SPS method 
against the Metropolis algorithm and found a much smaller dynamical 
scaling exponent for the SPS method.

To implement the SPS method, we break up the total potential 
$V = \sum_{i<j} u(r_{ij})$, where $r_{ij}$ is the distance
between particles $i$ and $j$, in two stages.  First, we 
can obviously break $V$ up 
into the individual pair interactions $u(r_{ij})$.  Next, for each of
the pair interactions, we can further decompose it into its
positive and negative parts, $u = u_+ + u_-$, such that 
$u_+(r)$ is everywhere zero except for $r<\sigma$ where $u_+(r) 
= u(r)$, and $u_-$ is its complement.  With this, the total potential 
becomes $V = \sum_{i<j} u_+(r_{ij}) + u_-(r_{ij})$.  
(We have also tried to decompose $u$ according to the WCA prescription 
\cite{83wee787} but found no major difference in the efficiencies of 
the two breakups.)  With this 
decomposition, $u_+$ is a purely repulsive potential, whereas $u_-$ is
bounded from above by zero.  
We apply the SPS algorithm to each of the $u_-(r_{ij})$ terms to try to 
switch it to $\tilde u_- = 0$ but keep all the $u_+(r_{ij})$ alive.
Since $u_-(r) \leq \tilde u_-$ for all $r$, $\Delta u_-^*$ can be 
simply set to 0.

The SPS algorithm, when applied to the Lennard-Jones fluid, proceeds 
as follows.  
Starting from the current configuration $\{\vec{r}_i\}$, 
we attempt to switch off each of the $u_-(r_{ij})$ one by one.
After the switching decision has been completed for every $u_-(r_{ij})$, the 
particles now interact with a modified potential in which some 
pairs of particles interact with $\bar u_-$ while the rest have 
zero attraction between them.
Since $u_+$ have been kept alive, every pair of particles also interact
through the purely repulsive $u_+$.  
At this point, one can employ any Monte Carlo move to update the system 
on this stochastically modified potential.  
One simple possibility is to apply a Metropolis algorithm with a local update 
to the SPS modified potential, just like on the original potential.
But as we will see, doing this alone will not improve 
the dynamical characteristics of the sampling.

\begin{figure}
\includegraphics[width=1.0\columnwidth]{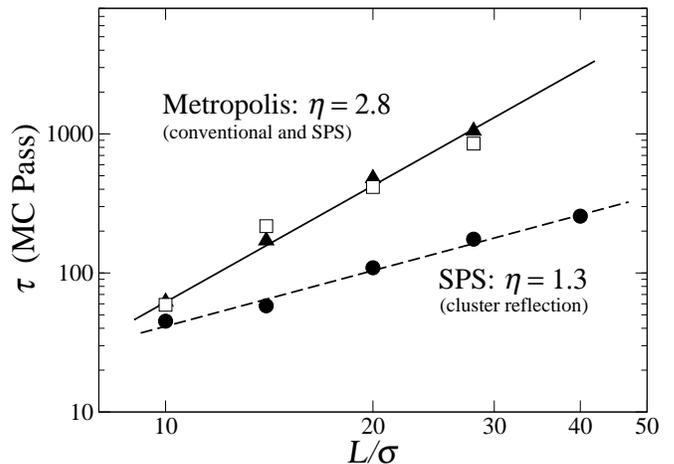}
\caption[]{
\label{fig:scaling}
Scaling of the 
autocorrelation time $\tau$ of the energy estimator with the box length $L$ 
in the simulation of a Lennard-Jones fluid at its critical point.
The dynamical exponents $\eta$ for the conventional Metropolis method 
on the original potential (squares) and on the SPS modified 
potential (triangles) are identical and equal approximately 2.8.  
The dynamical exponent in the SPS algorithm with the cluster 
reflection update (circles) is 1.3.
}
\end{figure}

The autocorrelation time $\tau$ in MC pass for the total energy measurement is 
shown in Fig.~\ref{fig:scaling} for different box sizes $L$, comparing the 
conventional Metropolis algorithm with a local update applied 
to the original potential (square) against 
the same Metropolis update applied to the SPS modified potential (triangles).
In both simulations, 
one MC pass is defined as having attempted one move for each particle 
in the system.  
Not surprisingly, the dynamic scaling behaviors of the two 
are identical to each other.
Since both simulations are based on the same local update method and they both 
satisfy detailed balance, the dynamical characteristics of the two sampling 
methods ought to be the same.
To improve sampling efficiency, nonlocal moves must be used.

In a near-critical system, the slowing-down problem is related to the 
divergence of the correlation length.  Update methods that 
employ only local moves will therefore have poor dynamical scaling, 
since they are unable to effect large-scale rearrangements 
of the system.  
SPS provides a basis for designing possibly more 
efficient alternative update schemes.
The system potential can be significantly 
simplified using SPS, enabling
large-scale moves to be performed with much higher acceptance ratio 
compared to the original potential.

\begin{figure}
\includegraphics[width=1.0\columnwidth]{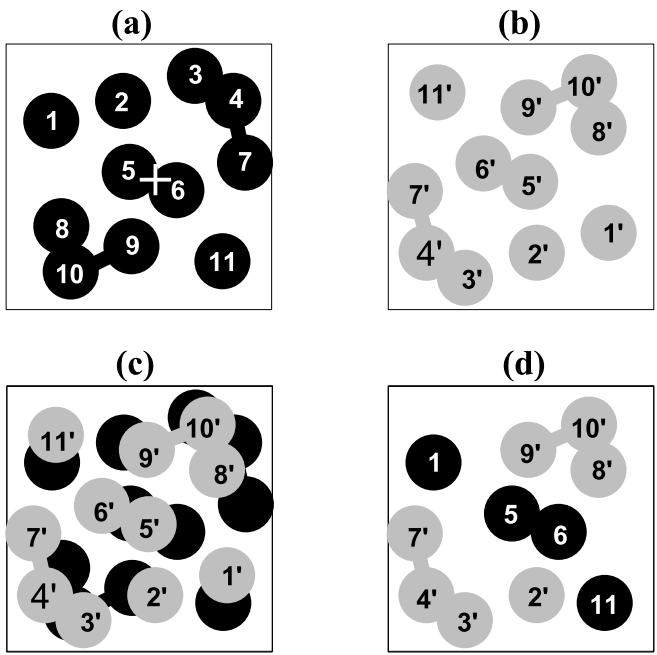}
\caption[]{
\label{fig:cluster}
Illustration of the cluster reflection algorithm.  
(a) Particles connected by frozen $\bar u_-$ 
(4-7 and 9-10) are bonded, shown in the figure connected by think lines.  
Particles that have a nonzero $u_+$ between them 
(3-4, 5-6 and 8-10) are shown as overlapping.  
Bonded and overlapping particles break up into disjoint clusters.  
In this example, there are six disjoint clusters.  This forms the 
``background'' configuration.  The cross indicates the position of 
the pivot, which for this illustration is near the center of the cell.
(b) All particles in the background are reflected across the pivot 
to generate the ``foreground'' configuration shown in grey.  The 
new position of each particle $i$ in the foreground is labeled $i^\prime$.
(c) Overlaying (b) on (a) generates 
superclusters from overlapping foreground and background clusters.  
In addition, foreground and background positions of the same particle are 
in the same cluster by default.
In this example, there are three disjoint superclusters:  
$(1^\prime,1,11^\prime,11)$, 
$(5^\prime,5,6^\prime,6)$, and a third encompassing the rest.
(d) For each supercluster, either all the foreground or 
all the background positions are accepted into the new configuration.  
In this example, two superclusters take on the background positions and
one supercluster takes on the foreground positions.
}
\end{figure}

To perform large-scale moves in the near-critical Lennard-Jones fluid 
on the SPS modified potential, we employed a simple scheme 
based on an algorithm originally proposed by Dress and Krauth \cite{95dre597} 
to treat hard sphere systems.  
We changed the method to suit the present situation, and 
our algorithm is illustrated in Fig.~\ref{fig:cluster} 
and proceeds as follows.
\begin{enumerate}
\item
For every pair of particles interacting with $\bar u_-$, we freeze 
their distance by placing a rigid ``bond'' between them.
\item
For every pair of particles that have a nonzero $u_+$ between them, 
we consider them to be ``overlapping'' and also freeze their distance.
\item
Based on the bonds and overlaps, we break the particles up into disjoint 
clusters.  Two clusters are disjoint if there is no bond or overlap 
between them.  This constitutes the ``background'' configuration in 
Fig.~\ref{fig:cluster}(a).
\item
To move the clusters, a point inside the simulation box is randomly selected
to be the pivot (illustrated by the cross in Fig.~\ref{fig:cluster}(a)), 
and all particles in the box are reflected across the pivot to obtain 
the ``foreground'' configuration in Fig.~\ref{fig:cluster}(b).  
\item
When the foreground is overlaid on the background, the clusters from 
the foreground and background form additional overlaps (but no additional 
bonds).  
The foreground and background positions of the same particle 
are also by default considered to be in the same cluster.  
Overlapping clusters can then 
be broken up into disjoint superclusters as in Fig.~\ref{fig:cluster}(c).
\item
Since there is no overlap between any particles 
(either foreground or background) 
from two disjoint superclusters, we can choose randomly to accept either the 
foreground or background positions inside each supercluster with equal 
probability.  An example of the resulting new configuration is illustrated
in Fig.~\ref{fig:cluster}(d).
\end{enumerate}
This completes one pass in our SPS simulation.  At this point, the 
simulation can start over with another cluster move from step 1, or 
we can carry out a Metropolis sweep before going back to 1.  Notice
that since the switching is done stochastically, a different cluster
structure would be generated every time even if the switching is applied 
to the same configuration.

In is easy to show that the cluster reflection algorithm above satisfies 
detailed balance in a trivial way, because the move conserves the 
nonzero part of the total potential of the system by fixing all the 
bonds and overlaps and the reflection clearly produces symmetric 
transition probabilities.  However, by itself the cluster reflection 
algorithm is nonergodic, because particles that have no overlap with each 
other in the configuration before the move will have no overlap either 
after the move.  
To have an ergodic Monte Carlo simulation, this cluster reflection algorithm
must be mixed with another ergodic move, such as Metropolis 
using a local update.  In our simulations, we performed one Metropolis move for
every 10 cluster updates, which adds minimal costs to the CPU time.

Autocorrelation times $\tau$ in MC pass for the total energy measurement is 
shown in Fig.~\ref{fig:scaling} for the 
SPS algorithm using the cluster reflection update (circles).  
For the SPS algorithm, one MC pass is defined as having
made one cluster reflection move plus one-tenth of a Metropolis move 
(needed for ergodicity).  In actual CPU time, a SPS MC pass is 
about 20\% {\em faster} than a Metropolis MC pass.
Near the critical point, the autocorrelation time is expected to 
scale with system size as $\tau \sim L^\eta$, and the 
dynamical scaling exponent $\eta$ is a measure of the efficiency of 
the MC method.  Clearly, the SPS method has a much smaller 
dynamical exponent.  In terms of absolute efficiency, the SPS algorithm
is more than ten times better for the largest simulations considered 
($L/\sigma = 40$ with 20464 particles), and accounting for CPU time 
difference, the SPS algorithm is actually 13 times better.

A generalized 
geometric cluster algorithm that is also based on the Dress and Krauth 
cluster move has also been proposed recently by Liu and Luijten 
\cite{04liu035504}.  In their approach, the energies 
of the configuration before and after the cluster move have to be computed
to determine the transition probabilities; whereas in our SPS algorithm, 
only the energy {\em before} the move has to be computed in order to 
determine the switching probabilities.  In cases where the potential 
is complicated and costly to compute, the SPS algorithm here will offer 
CPU time savings compared to the method of Liu and Luijten, but it may 
suffer from lower switching rates.

\section{Conclusions} \label{sect:conclusions}

In summary, we have presented a new Monte Carlo method that 
is based on a stochastic potential switching algorithm.  This new 
algorithm enables the equilibrium 
properties of a system with potential $V$ to be computed using a 
Monte Carlo simulation for a system with a possibly less complex 
stochastically altered potential $\tilde V$.  
Generalization of this method to systems with additive potentials 
provides for an efficient scheme for simulating complex systems.   
The validity of the method is illustrated with a simple one-dimensional 
example, and its practical utility in alleviating the critical slowing-down 
problem is illustrated with a Lennard-Jones fluid near its critical point.

\begin{acknowledgments}

This work was supported by the National Science Foundation under grant 
CHE-9970766. 

\end{acknowledgments}

\end{document}